# Tunable Photodetectors via in situ Thermal Conversion of TiS$_3$ to TiO$_2$

Foad Ghasemi [1,2,†], Riccardo Frisenda [3,†,*], Eduardo Flores [4], Nikos Papadopoulos [5], Robert Biele [6,7], David Perez de Lara [1], Herre S. J. van der Zant [5], Kenji Watanabe [8], Takashi Taniguchi [8], Roberto D'Agosta [6,9], Jose R. Ares [4], Carlos Sánchez [4,10], Isabel J. Ferrer [4,10] and Andres Castellanos-Gomez [3,*]

[1]  Instituto Madrileño de Estudios Avanzados en Nanociencia (IMDEA-Nanociencia), Campus de Cantoblanco, E-28049 Madrid, Spain
[2]  Nanoscale Physics Device Lab (NPDL), Department of Physics, University of Kurdistan, 66177-15175 Sanandaj, Iran
[3]  Materials Science Factory, Instituto de Ciencia de Materiales de Madrid (ICMM-CSIC), E-28049, Madrid, Spain.
[4]  Materials of Interest in Renewable Energies Group (MIRE Group), Dpto. de Física de Materiales, Universidad Autónoma de Madrid, UAM, Campus de Cantoblanco, E-28049 Madrid, Spain
[5]  Kavli Institute of Nanoscience, Delft University of Technology, Lorentzweg 1, Delft 2628 CJ, The Netherlands
[6]  Nano-Bio Spectroscopy Group and European Theoretical Spectroscopy Facility (ETSF), Universidad del País Vasco CFM CSIC-UPV/EHU-MPC & DIPC, Av.Tolosa 72 ,20018, San Sebastián, Spain
[7]  Institute for Materials Science and Max Bergmann Center of Biomaterials, TU Dresden, 01062 Dresden, Germany
[8]  National Institute for Materials Science, Namiki 1-1, Tsukuba, Ibaraki 305-0044, Japan
[9]  IKERBASQUE, Basque Foundation for Science, 48013 Bilbao, Spain
[10] Instituto Nicolás Cabrera, Universidad Autónoma de Madrid, UAM, Campus de Cantoblanco, E-28049 Madrid, Spain.
*   Correspondence: riccardo.frisenda@csic.es (R.F.); andres.castellanos@csic.es (A.C.-G.)
†   These authors contributed equally to this work

**Abstract:** In two-dimensional materials research, oxidation is usually considered as a common source for the degradation of electronic and optoelectronic devices or even device failure. However, in some cases a controlled oxidation can open the possibility to widely tune the band structure of 2D materials. In particular, we demonstrate the controlled oxidation of titanium trisulfide (TiS$_3$), a layered semiconductor that has attracted much attention recently thanks to its quasi-1D electronic and optoelectronic properties and its direct bandgap of 1.1 eV. Heating TiS$_3$ in air above 300 °C gradually converts it into TiO$_2$, a semiconductor with a wide bandgap of 3.2 eV with applications in photo-electrochemistry and catalysis. In this work, we investigate the controlled thermal oxidation of individual TiS$_3$ nanoribbons and its influence on the optoelectronic properties of TiS$_3$-based photodetectors. We observe a step-wise change in the cut-off wavelength from its pristine value ~1000 nm to 450 nm after subjecting the TiS$_3$ devices to subsequent thermal treatment cycles. Ab-initio and many-body calculations confirm an increase in the bandgap of titanium oxysulfide (TiO$_{2-x}$S$_x$) when increasing the amount of oxygen and reducing the amount of sulfur.

**1. Introduction**

Low-dimensional semiconductors are attracting increasing interest in the scientific community working on optoelectronic devices thanks to their outstanding optical and electronic properties combined with reduced dimensionality [1–3]. The large surface-to-volume ratio of two-dimensional (2D) materials benefits many applications such as gas-sensing, but it may enhance the sensitivity of these materials to oxidation compared to bulk materials. Moreover, lattice vacancies and atomic-level defect combined with the presence of light can accelerate the oxidation process [4–8], which is typically accompanied by a degradation of the electrical and optical properties reducing the device performance [9–10]. Furthermore, shining high intensity light on 2D materials can induce additional processes of photo-oxidation [11–14]. The overall performance reduction induced by oxidation seems to be one of the main issues to solve in developing industrial applications based on 2D materials, therefore controlling the oxidation process is a very active subject for both fundamental and applied research in the context of band engineering.

Titanium trisulfide ($TiS_3$) is a layered semiconductor which has attracted much attention recently thanks to its quasi-1D electronic and optoelectronic properties [15–18] and its direct bandgap of 1.1 eV [19–25]. Using first-principles calculations, Iyikanat et al. showed that $TiS_3$ can react with various forms of oxygen [26]. An experimental demonstration was given by Molina-Mendoza et al., who reported thermogravimetric analysis (TGA) of bulk $TiS_3$ in oxygen atmosphere showing the partial conversion of the material into $TiO_2$, a large bandgap (3.2 eV) insulator with a wide range of applications [19,27–29]. In this article we investigate the controlled thermal oxidation of individual $TiS_3$ nanoribbons and its influence on the optoelectronic properties of $TiS_3$-based photodetectors. We first study the oxidation of $TiS_3$ powder and single nanoribbons deposited on a glass substrate. Using Raman spectroscopy and optical analysis we can monitor the material properties as a function of time while heating at 320 °C in air. We find that an individual $TiS_3$ nanoribbon converts to crystalline $TiO_2$ in approximately 10 minutes. Control experiments performed on $TiS_3$ nanoribbons fully encapsulated between hexagonal boron nitride flakes confirm that the direct contact between $TiS_3$ and air is necessary for the oxidation process to happen. After establishing the change in material properties we demonstrate the controlled oxidation of a $TiS_3$ nanoribbon photodetector that allows tuning the cut-off wavelength and sensitivity of the device. By monitoring the change in its current–voltage characteristics and in its spectral photoresponse, we find that the cut-off wavelength is blue-shifted upon oxidation, reaching a cut-off wavelength of 450 nm (while it is ≈1100 nm for pristine $TiS_3$ according to previous works [19]). Various intermediate states are observed, demonstrating the tunability of the nanoribbon bandgap. Ab-initio and many-body calculations confirm an increase in the bandgap near to that of titanium oxysulfide ($TiO_{2-x}S_x$) when increasing the amount of oxygen and reducing the amount of sulfur.

**2. Materials Synthesis**

The starting $TiS_3$ material was synthetized by a solid-gas reaction using Ti powder and sulfur powder sealed into a quartz ampoule and kept at 550 °C for 20 h. Additional details about the synthesis and elemental characterization of $TiS_3$ can be found in the references [17,30,31]. Figure 1a shows an optical picture of $TiS_3$ powder while Figure 1b shows the same powder after heating it for 5 min at 350 °C in air. After heating the material, we observed a dramatic change of its appearance, with a clear color from black to white, due to the conversion of $TiS_3$ to $TiO_2$. Thanks to the layered structure of $TiS_3$, individual nanoribbons can be isolated by mechanical exfoliation. To study an individual $TiS_3$ nanoribbon, we first exfoliated the powder onto Nitto tape and then transferred part of the flakes from the tape to a viscoelastic polidimethylsiloxane (PDMS) stamp. After the identification, we transferred the chosen nanoribbon to a different substrate (such as glass, $SiO_2$/Si…) with an all-dry deterministic transfer method [32,33].

A typical TiS₃ nanoribbon transferred onto a glass slide is shown in Figure 1c. The microscope picture is recorded in transmission mode and the nanoribbon appears black since it is absorbing most of the white light due to the bandgap of 1.1 eV. The nanoribbon has a length of approximately 150 μm (oriented along the crystal *b* axis) and a width of 5 μm (*a* axis). Figure 1d shows the same nanoribbon after heating it for 5 min at 350 °C in air. The morphology of the nanoribbon appears intact, but the substantial change in color indicates that its absorption, which is related to the bandgap and the band-structure, has changed dramatically. The change from black to white/transparent due to a reduction of the light absorption indicates an opening of the bandgap. A higher resolution picture of the initial and final status of TiS₃ nanoribbons can be obtained using electron microscopy. Figure 1e-f shows a field emission gun scanning electron microscopy (FEGSEM) image of TiS₃ nanoribbons before and after the heat treatment. As can be seen from the images, the nanoribbons morphology is maintained after the heat treatment, although the final nanoribbons show higher roughness.

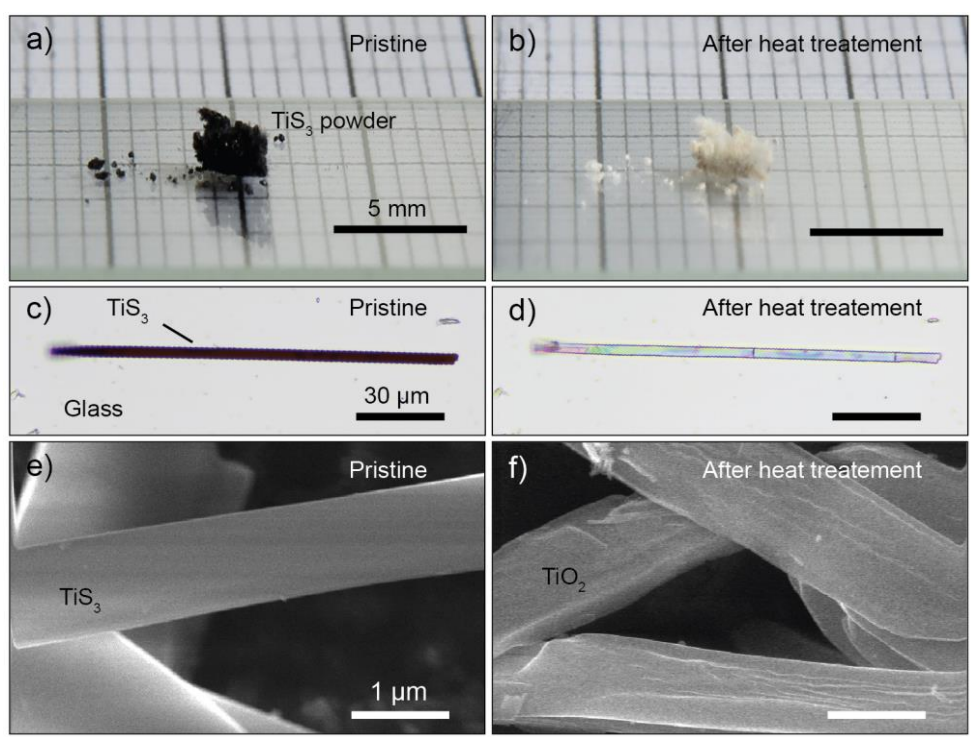

**Figure 1. a**) Photograph of titanium trisulfide (TiS₃) powder onto a glass slide. **b**) Photograph of the same powder of panel (a) after heating in air for 5 min at 350 °C. **c**) Optical image of an individual TiS₃ nanoribbon transferred onto a glass substrate, recorded in transmission illumination mode under the microscope. **d**) Same as (c) after heating in air for 5 min at 350 °C. **e–f**) Field emission gun scanning electron microscopy (FEGSEM) image of pristine TiS₃ (e) and after heating at 400 °C in air (f). Note that the images in panel (e) and (f) correspond to different nanoribbons.

## 3. Experimental Results and Discussion

*3.1 Thermal Oxidation Analysis*

To gain deeper insight into the thermal oxidation process of TiS₃ we used TGA coupled to mass spectrometry (MS). Figure 2a shows a TGA curve of TiS₃ kept under a flux of 90 mL/min of air and heated at a rate of 10 °C/min. The graph of temperature as a function of time, shown in the top panel of Figure 2a, displays a discontinuity at approximately 30 min, indicative of an exothermic reaction occurring at 300 ± 10 °C. This reaction is accompanied by a loss of approximately 43% of the initial mass (see the bottom panel of Figure 2a) that is consistent with the difference between the mass of TiS$_3$ and TiO$_2$ (44%). These results indicate that the conversion of TiS$_3$ into TiO$_2$ takes place under atmospheric conditions starting at 300 °C. A more in-depth look can be achieved using a mass spectrometer to detect the species present during the reaction. Figure 2b shows the ionic currents at $m/q$ = 16, 32 and 48 ($m/q$ is the ration between the atomic mass $m$ and the atomic charge $q$ of the species) as a function of time recorded during the TGA experiment, corresponding to O$_2$ (whose cracking pattern shows two signals) and SO$_2$. The dip observed at 30 minutes in the traces of $m/q$ = 16 and 32 indicated that the O$_2$ present in the atmosphere was reacting with TiS$_3$. At the same time, the peak in current of $m/q$ = 48 is consistent with the liberation of sulfur atoms from TiS$_3$ and their successive reaction with oxygen to form gaseous SO$_2$.

The composition of the final product after the thermal treatment of TiS$_3$ was studied with x-ray diffraction measurements (XRD). The x-ray diffraction measurements were performed using a X-pert PRO diffractometer under a θ/2 θ configuration. Figure 2c shows XRD patterns of TiS$_3$ before and after the heating treatment with the main diffraction planes indicated. In the pristine material a single polycrystalline phase is observed, TiS$_3$. All the diffraction peaks in the XRD pattern can be indexed to the phase monoclinic TiS$_3$ in good agreement with the standard JCPDS card no. 00-036-1337. The most intense diffraction peaks at Bragg angles 10.2°, 20.4°, 30.8° and 41.4°are assigned to the (001), (002), (003), (012) and (004) planes. After the heating, the XRD pattern changed significantly. The new diffraction peaks can be indexed to polycrystalline tetragonal TiO$_2$, anatase, whose more intense peaks at angles 25.3°, 38.6°, 37.8°and 48.1° can be attributed to the (101), (004), (112) and (200) planes in good agreement with the standard JCPDS card no. 00-021-1272. Comparing the two spectra it can be seen that the diffraction peaks after the treatment are broader than the ones before, indicating that the produced TiO$_2$ has crystallites of a smaller size. Overall, the previous results support a scenario in which TiS$_3$ heated above 300 °C in ambient conditions undergoes the global exothermic reaction: TiS$_3$ + 4O$_2$ → TiO$_2$ + 3SO$_2$, which converts the trisulfide in anatase TiO$_2$.

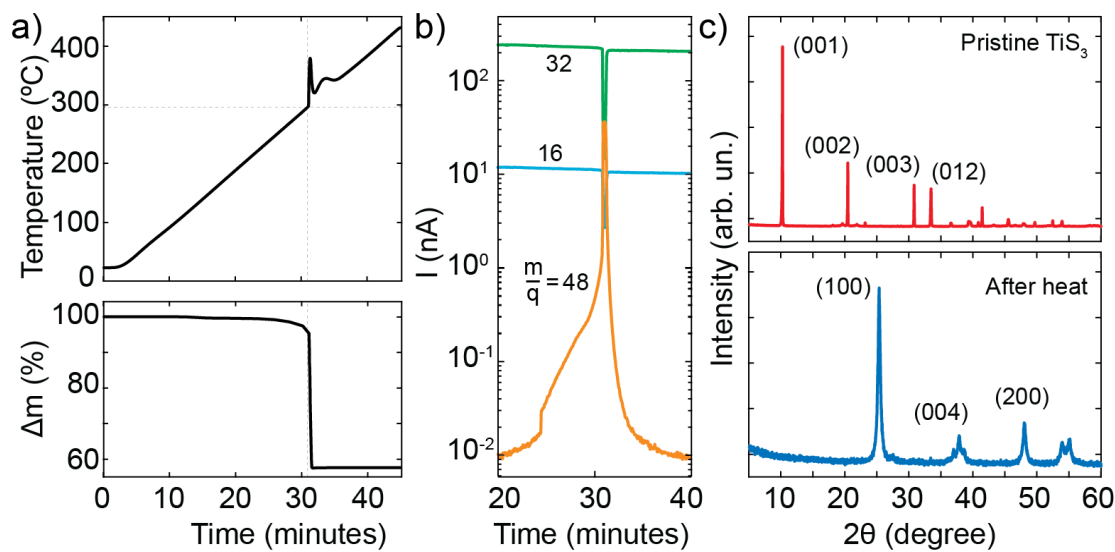

**Figure 2. a**) Thermogravimetric analysis (TGA) curves of TiS$_3$ during the heating under a flux of air of 90 mL/min at 10 °C/min, time dependence of the temperature (top) and time evolution of the loss of mass (bottom). **b**) Ionic currents at different m/q ratios as a function of time during the TGA experiment. **c**) X-ray diffraction (XRD) patterns before and after the heating treatment.

After characterizing the thermal oxidation of bulk TiS$_3$, we focused on individual nanoribbon oxidation which was interesting for the fabrication of high-quality optoelectronic devices. We started by using Raman spectroscopy to study the composition of a single nanoribbon. Figure 3a shows the Raman spectra of a TiS$_3$ nanoribbon recorded in its pristine form (after deposition onto a SiO$_2$/Si substrate) and during a heating cycle with spectra taken every 2 minutes while heating the sample at a temperature of 320 °C. The Raman spectra were recorded in a Raman Microscope (SENTERRA II, Bruker) while illuminating the sample with a laser of 532 nm focused in a circular spot (area ~2 µm$^2$, power 2 mW, power density 1 mW/µm$^2$) and an integration time of 20 s. The power density that we use is lower than the threshold density for photooxidation of the TiS$_3$, which we estimate to be 5 mW/µm$^2$ (see Section S4 of the Supporting Information). The Raman signal of the pristine TiS$_3$ shows four prominent peaks due to TiS$_3$ and a very weak peak at 520 cm$^{-1}$ due to the silicon substrate. The peaks at energies 177 cm$^{-1}$, 302 cm$^{-1}$, 371 cm$^{-1}$, and 559 cm$^{-1}$ correspond to A$_g$ Raman modes of the TiS$_3$ nanoribbon and are in good agreement with the modes reported for bulk TiS$_3$ [34]. After heating up the sample we observe a reduction in the intensity of the TiS$_3$ peaks and an increase in the Si peak intensity during the first two cycles (4 min) that can be attributed to an increase in the transparency of the nanoribbon. After approximately 6 minutes of heating (after three cycles) we observed the quenching of the TiS$_3$ peaks with only the 520 cm$^{-1}$ Si peak visible in the Raman spectrum of the sample. The spectra recorded after 8, 10 and 12 minutes of heating show the appearance of a new peak centered at 142 cm$^{-1}$. This peak is consistent with the signature of an *E$_g$* Raman mode of TiO$_2$ [35,36]. The evolution of the Raman spectra of the nanoribbon shows that a pristine TiS$_3$ nanoribbon can be converted to TiO$_2$ by heating at 320 °C. The TiS$_3$ to TiO$_2$ oxidation process can be readily visualized from the plot in Figure 3b in which we show the extracted intensities of the 142 cm$^{-1}$ (TiO$_2$) and 302 cm$^{-1}$ (TiS$_3$) peaks as a function of the number of heating cycle. Apart from the change in the Raman signal, we also observe a clear change in the color of the nanoribbon deposited on the SiO$_2$/Si surface from green to yellow during the conversion process as shown in the inset of Figure 3b.

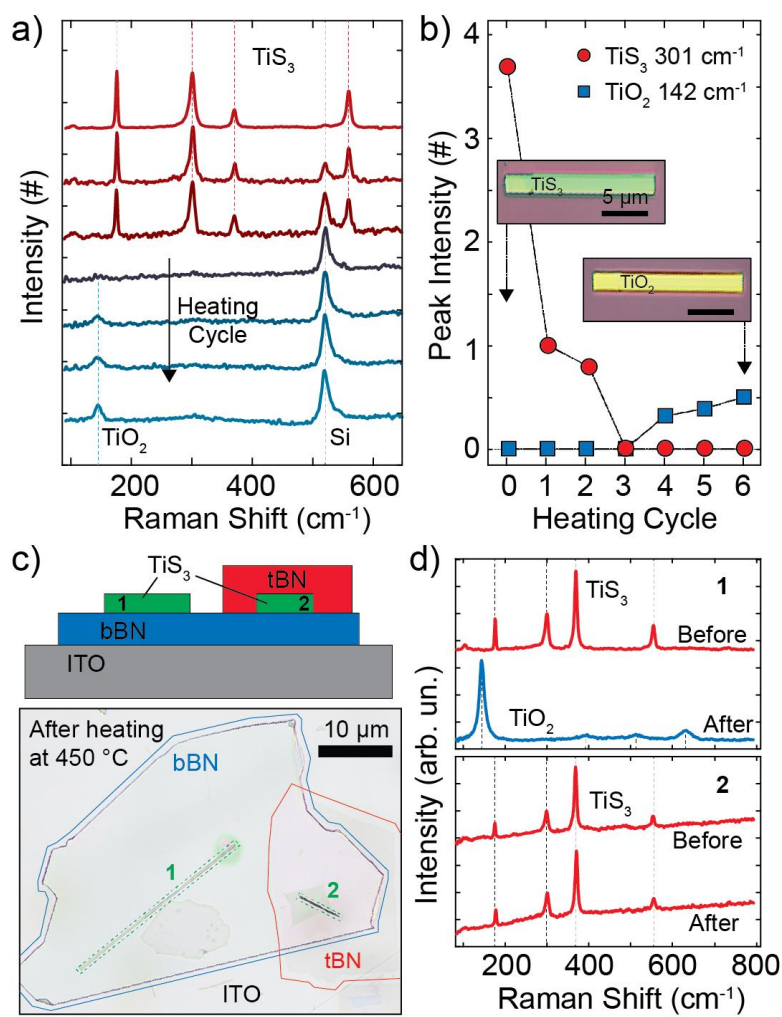

**Figure 3. a)** Raman spectra of a TiS$_3$ nanoribbon onto a SiO$_2$/Si substrate recorded in its pristine state (top) and during heating cycles at 320 °C. The spectra have been offset for clarity and each cycle corresponds to 2 min at 320 °C. **b)** Intensity of the peaks at 142 cm$^{-1}$ and 301 cm$^{-1}$ as a function of heating cycle. The inset shows an optical picture of the pristine TiS$_3$ nanoribbon onto SiO$_2$/Si (left) and of the same nanoribbon after heat treatment (right). **c)** Schematic of the boron-nitride/TiS$_3$ stack (top) where 1 is not encapsulated and 2 is fully encapsulated. Optical picture of the sample after heating it at 450 °C for 30 min (bottom). We highlighted the contour of two boron nitride flakes (in red and blue) and of the nanoribbons (green) for clarity. **d)** Raman spectra of the unencapsulated (1, top) and encapsulated (2, bottom) nanoribbons of panel (c) recorded before and after heating the sample.

To study the role of the environment in the oxidation process of a single nanoribbon we fabricated a hexagonal boron nitride (h-BN) encapsulated TiS$_3$ nanoribbon on top of a transparent indium tin oxide (ITO) substrate. The top panel of Figure 3c shows the schematic of the samples. We first transferred a flake of h-BN onto the ITO surface and then transferred two TiS$_3$ nanoribbons onto the h-BN surface. We finally transferred a second h-BN flake covering just one of the two TiS$_3$ nanoribbons. Figure 3c shows an optical picture of the fabricated stack after heating it. We recorded the Raman spectrum of each nanoribbon before and after heating up the sample at 450 °C (a temperature much larger than the threshold for oxidation of 300 °C). Figure 3d shows the Raman spectra of the two nanoribbons before heating that display very similar features and are both characterized by the four TiS$_3$ peaks discussed above. Notice that compared to Figure 3a the Si peak is missing since the substrate is ITO. When comparing the spectra after heating at 450 °C for 30 min we observe a large difference between the two nanoribbons. While the fully encapsulated nanoribbon (2) does not show a significant change in its spectrum, indicating that the final material is TiS$_3$, the unencapsulated flake (1) shows a dramatic change in its spectrum, due to the conversion from TiS$_3$ to TiO$_2$. The encapsulation (with bottom and top h-BN) prevents the oxidation of the TiS$_3$ nanoribbons from happening. The effect of the h-BN layers are visible also in the optical picture of Figure 3c where a big difference in the aspect of the two nanoribbons (one being oxidized and the other not) can be seen. As a side note, the Raman spectrum of the oxidized uncovered nanoribbon shows additional peaks compared to Figure 3a, located at energies of 396 cm$^{-1}$, 518 cm$^{-1}$ and 635 cm$^{-1}$. These peaks, associated to the TiO$_2$ crystalline phase, are due to anatase TiO$_2$ in agreement with the results from the XRD measurements of Figure 2c.

*3.2. Bandgap Energy Calculation*

The experiments discussed above show that a thermal oxidation process can convert TiS$_3$ nanoribbons into TiO$_2$ and that this evolution can be followed on a single ribbon level with Raman spectroscopy. We now focus on the changes in the band-structure of the system when passing from TiS$_3$ to TiO$_2$.

To calculate the electronic band structure, we have performed state-of-the-art ab-initio Density Functional Theory (DFT) calculations with a pseudo-potential plane-wave method as implemented in the PWSCF code of the Quantum-ESPRESSO suite [37–39]. Figure 4 shows some examples of the electronic band structure calculated with DFT for the TiO$_2$ in the rutile, anatase, IV and V allotropes (see a 3D representation of the different structures in Figure 5a) over the first Brillouin zone. A GW calculation most of the time opens up the band gap with a rigid shift of the conduction bands.

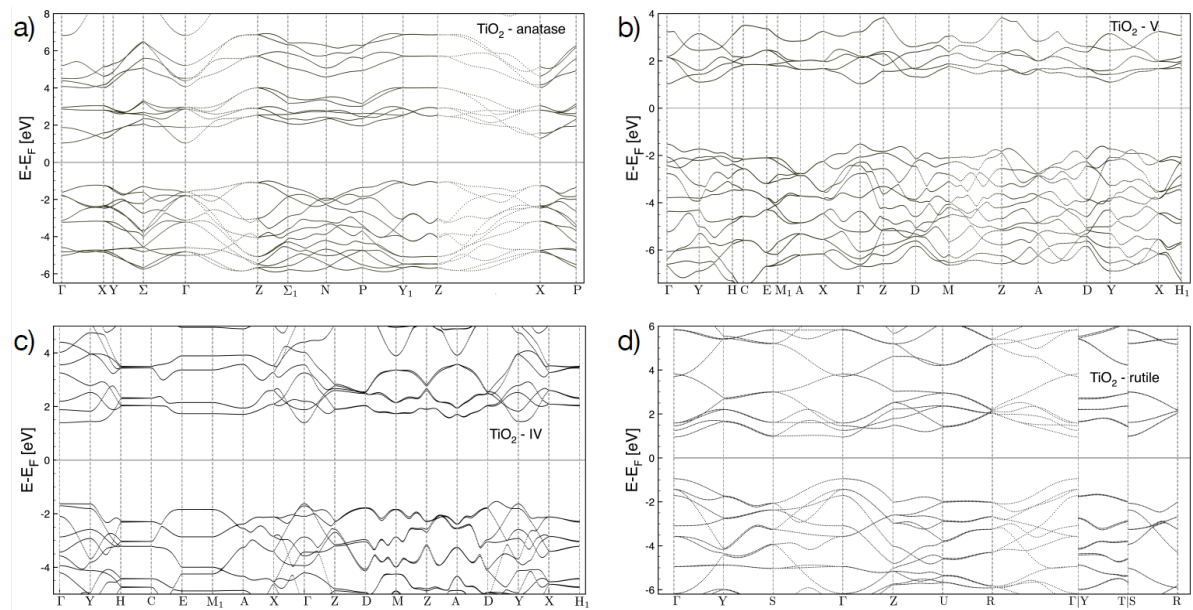

**Figure 4.** Band structure calculated with Density Functional Theory (DFT) along the first Brillouin zone for the different allo-tropes of the TiO$_2$, (**a**) anatase, (**b**) V, (**c**) IV, and (**d**) rutile.

For both Ti and S, the electron exchange-correlation potential is evaluated within the generalized gradient approximation throughout the Perdew–Burke–Ernzerhof's functional. For S the Martins–Troulliers', while for Ti the Goedecker–Hartwigsen–Hutter–Teter's pseudo-potentials are used, including semi-core states for the valence electrons. In all structures we have optimized the atomic positions with a residual force after relaxation of 0.001 a.u. and have also included van der Waals corrections. The kinetic energy cut-off for the plane wave basis set is at 220 Ry, while the cut-off for the charge density is 880 Ry. The sampling of the Brillouin zone is 6x6x6 according to the Monkhorst–Pack scheme. The parameters chosen ensure a convergence of the DFT band gap within an accuracy of around 0.01 eV. In general, DFT underestimates the band gap, however those results might be used to estimates tendencies, like an increase or decrease in the band gap under oxidation. To enable a better comparison with the experimental values, we have further performed a more refined calculation for some of the structures based on non-self-consistent GW method. This opens up the DFT gap (0.37 eV) for pristine $TiS_3$ to the experimental levels (about 1.2 eV). The GW band gaps have been converged within an accuracy of around 0.05 eV.

In order to construct the intermediate oxidation structures, we have started with the relaxed structures of pristine $TiS_3$ and have replaced three S atoms (in a unit-cell of eight atoms) with two O atoms. We have relaxed possible oxidation states by varying the position of the atom replacements. The three structures in Figure 5a correspond to the ones of the lowest total energy, which are most likely to be formed during the oxidation process. Similar techniques have been applied to find the structures for the fully oxidized states. For these structures, we have also performed a GW calculation to evaluate the band gap. Both DFT and GW show an increase in the gap at the $\Gamma$-point of the band structures of interest for these experiments, ranging from about 1eV for $TiS_3$ to about 3 eV for the $TiO_2$ in the rutile structure. For intermediate oxidation states, the evaluation of the actual atomic configuration is more difficult since one should consider exceedingly large super cells to build the possible atomic configurations, but our results show clearly an increase in the DFT band gap from 0.31 eV for $TiS_3$ to about 0.7 eV for the partially oxides structures, to more than 2 eV for the $TiO_2$ in the different allotropic forms. These results are consistent with others already present in the literature for $TiO_2$.

Figure 5a shows the calculated bandgap of $TiS_3$ (left), intermediate $Ti_2S_3O_2$ phases (middle) and of four different polytypes of $TiO_2$ (right). The different materials are ordered along the horizontal axis according to the total energy (when going from left to right the total energy decreases and the thermodynamic stability increases). Because DFT typically underestimates the band gap energy [40], we have further performed a more refined calculation for some of the structures based on non-self-consistent GW method.

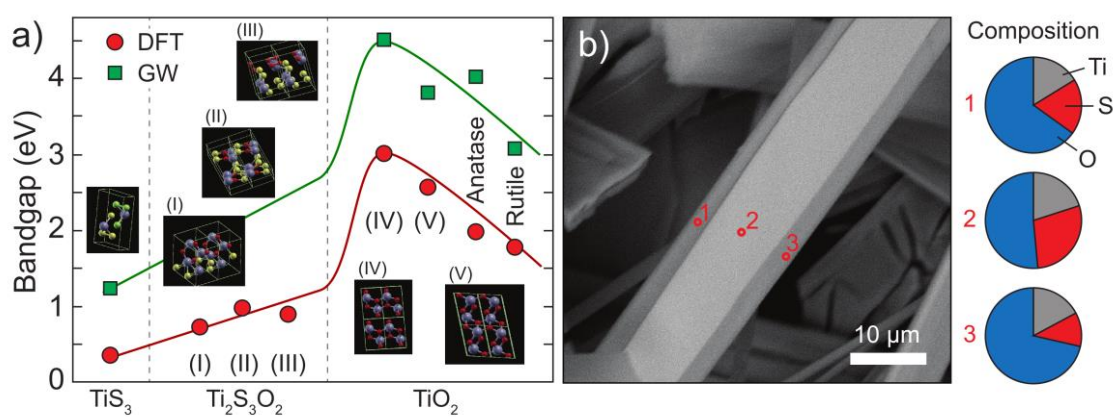

**Figure 5. a**) Bandgap calculated from DFT (red circles) and GW (green squares) for $TiS_3$, $TiO_2$ and intermediate phases composed of 50% $TiS_3$ and 50% of $TiO_2$ (indicated as $Ti_2S_3O_2$). The lines are guides-to-the-eye. **b**) FEGSEM image of a $TiS_3$ nanoribbon heated up to 300 °C during 1 h. Spatially resolved electron diffraction measurements at the positions indicated by the red dots (right panel) reveal a higher oxygen concentration along the ribbon edges.

Both DFT and GW calculations predict that the bandgap energy increases when the content of sulfur decreases and the oxygen increases. The stable intermediate titanium oxysulfide phases predicted by the theory are consistent with FEGSEM measurements of partially oxidized nanoribbons.

The left panel of Figure 5b shows a FEGSEM image of such a nanoribbon in which the core and the edges show a different contrast. The Energy-dispersive X-ray spectroscopy (EDX) analysis of the image performed in the center and at the edges of the nanoribbon reveals a higher oxygen concentration along the edges. This indicates that the oxidation process occurs through the formation of intermediate phases $TiO_{2-x}S_x$ at the ribbon surface that form a sheath around $TiS_3$. This phenomenon is also visible in the microscope pictures of partially oxidized nanoribbons in Figure S2, Section S1 of the Supporting Information.

*3.3. Electronic and Optoelectronic Characterization*

After the change in the optical appearance and vibrational properties of $TiS_3$ nanoribbons after high temperature treatment due to the oxidation of $TiS_3$ to $TiO_2$ was established, we then investigated electronic transport through an individual nanoribbon and monitored the change of the optoelectronic properties. We fabricated a $TiS_3$ photodetector by transferring an individual $TiS_3$ nanoribbon onto two pre-patterned platinum electrodes separated by a distance of 20 μm. The electrical measurements were carried out in atmospheric conditions using a home-built probestation equipped with a source measure unit (Keithley 2450). For the optoelectronic measurements the devices were illuminated by focusing the light of different high-power fiber-coupled LED sources with different emission wavelengths (Thorlabs), forming a circular spot (400 μm in diameter) onto the surface of the sample. The total optical power reaching the sample was measured with a silicon photodetector (Thorlabs power meter PM100D with sensor S120VC).

Figure 6a shows a schematic of the device and an optical image of the pristine $TiS_3$ device. The $TiS_3$ nanoribbon bridges the two electrodes and light can be shined on the exposed $TiS_3$ channel to study its photoresponse. Figure 6b shows the current–voltage characteristics (*I-V*s) of the device recorded just after the fabrication (top panel) and after heating it for 12 min at 320 °C (bottom panel). From previous measurements the electrical resistivity of the exfoliated $TiS_3$ thin nanoribbons is ~0.1 Ω·cm [15] while measurements on macroscopic $TiS_3$ whiskers, reported by Gorlova et al., reach 2 Ω·cm[41]. Photographs of the device at various stages of the evolution are shown in Figure S6, Section S3 of the Supporting Information. The *I-V*s were recorded in dark conditions (black curve) and under global illumination at 405 nm with power density 1 W/cm² (purple curve). Comparing the *I-V*s of the pristine with those of the oxidized device we can see a difference in the shape of both *I-V*s and in the magnitude of the current. The $TiS_3$ device is characterized by linear *I-V*s and current in the range of μA while the oxidized device has non-linear *I-V*s with currents in the pA (approximately six orders of magnitude lower than the pristine device). In both cases the device responds to light, evidenced by the larger current observed in the *I-V*s under illumination at 405 nm in comparison to the ones recorded in the dark. Considering that the area of the channel is 100 μm² and that the incident optical power is 1 W/cm², the responsivity of the device to 405 nm at 3 V (−3 V) is 8 A/W (11 A/W) in the pristine ($TiS_3$) case and 0. 16 mA/W (0. 22 mA/W) in the oxidized case. In total in this work we fabricated and characterized 11 devices that have been heated at 320 °C in air, see Section S3 of the Supporting Information. In four cases out of 11 we observed the conversion of the $TiS_3$ photodetector in a $TiO_2$ one without losing the functionality (success rate 36%).

In order to study the responsivity of the device to different wavelengths and incident optical powers we measure current–time traces (at fixed bias voltage) while switching ON and OFF in time with the incident light. Figure 6c shows various photocurrents versus time traces measured with wavelengths in the range between 375 nm and 660 nm. At the beginning of the measurement the light is switched OFF and the current passing through the device has only the dark current contribution. When switching ON the illumination (approximately at time = 5 s in the plot), the current passing through the device rapidly increases thanks to the additional contribution given by the photogenerated current. By extracting the current difference between the OFF and ON illumination, we can calculate the photocurrent and responsivity of our device. From the plot in Figure 6c one can see that the responsivity of the TiS$_3$ pristine device decreases when increasing the wavelength of the incident radiation. Figure 6d shows the photocurrent of the pristine device at 405 nm for different values of the illumination power density showing larger values for higher incident powers. Figure 6e–f shows similar measurements to those shown in Figure 6c–d made on the oxidized nanoribbon photodetector. These measurements show that after oxidation the photocurrent decreases and the time response increases. Moreover, the photodetector responds only to light with a wavelength shorter than 405 nm. The cut-off wavelength of the photodetector (defined as the largest wavelength for which the photodetector shows a response higher than the noise level 2 × 10$^{-6}$ A/W), is larger than 660 nm in the pristine case and blue-shifts to 405 nm in the oxidized state. As a control experiment we heated one device in high vacuum conditions (pressure ~10$^{-5}$ mbar) to test the effect of heat in the absence of oxygen, see Section S2 of the Supporting Information. This device maintained the photodetecting properties of TiS$_3$ (responsivity spectrum and cut-off wavelength) even after 30 min at 320 °C.

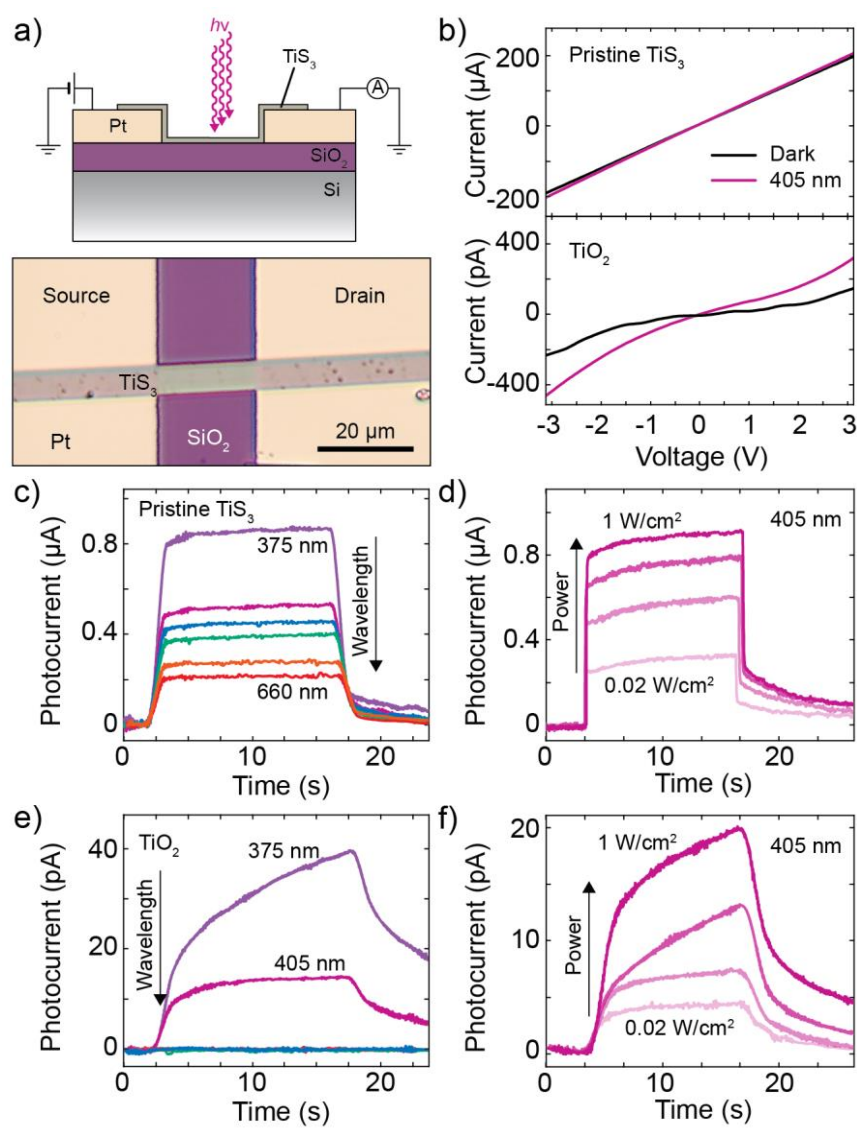

**Figure 6. a**) Schematic of a TiS$_3$ photodetector (top). Optical microscope image of a TiS$_3$ photodetector (bottom). **b**) Current–voltage characteristics of the sample in its pristine state (top) and after oxidation of the nanoribbon (bottom). The black line is the current recorded with the device kept in dark while the purple line is under illumination at 405 nm. (**c–f**) Photocurrent as a function of time recorded

while modulating the intensity of the incident light with a square wave on the pristine TiS$_3$ photodetector (c–d) and on the TiO$_2$ (e–f). The colors in (c,e) correspond to different incident wavelengths (375 nm, 405 nm, 420 nm, 530 nm, 605 nm, 660 nm) while the shades of purple in (d,f) correspond to different incident power densities from 0.02 W/cm$^2$ to 1 W/cm$^2$ at a fixed wavelength of 405 nm.

From current–time traces similar to the ones in Figure 6c–f we extracted the responsivity of the device, in its pristine state and after subsequent heating cycles, as a function of wavelength and incident power. Figure 7a shows the evolution of the wavelength-resolved responsivity of the device measured at an incident power of 100 nW during the heating process. The pristine device has a responsivity of approximately 1 A/W that after the first heating cycle decreases to 10$^{-4}$ A/W. The dependence on the wavelength also gets modified by the heating process. While in the first three traces the device responds to all the probed wavelengths from 375 nm to 660 nm, in the last four traces the responsivity shows an abrupt decrease for wavelengths larger than 450 nm, indicating that there is a blue-shift of the cut-off wavelength of the photodetector. The wavelength-resolved responsivity reported in the initial and final stages of the device are compatible with previously reported spectra of TiS$_3$ and TiO$_2$ photodetectors [17,42].

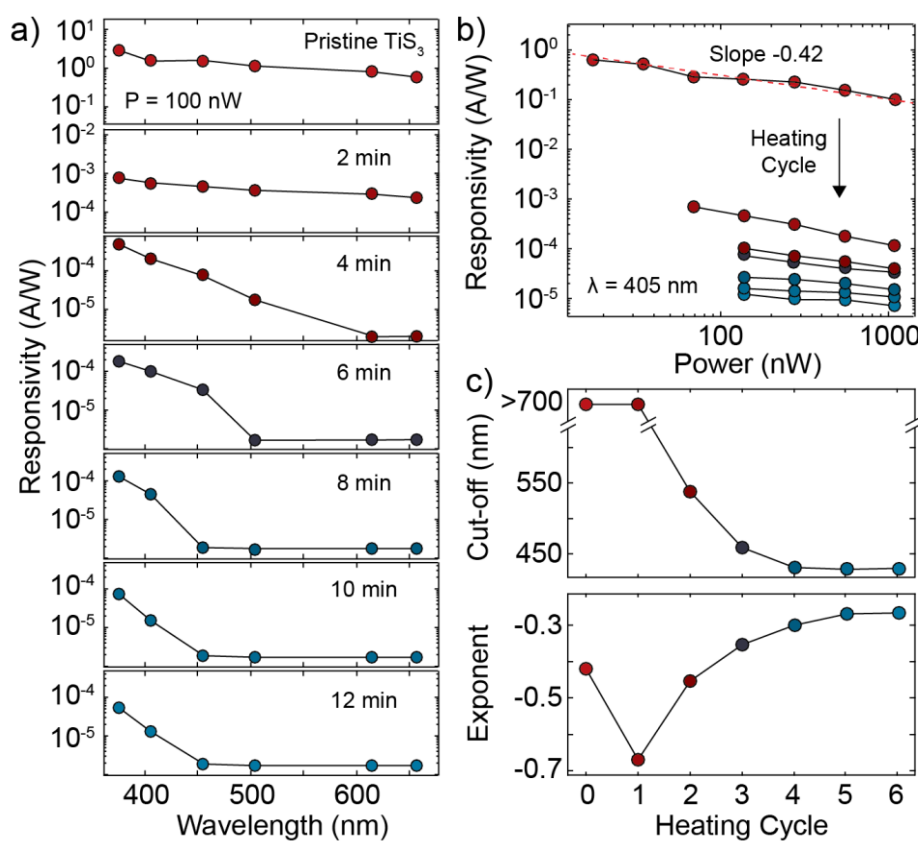

**Figure 7. a**) Responsivity of the device as a function of wavelength for different heating cycles. The pristine TiS$_3$ device (top) was heated at 320 °C in steps of 2 min and after each step the responsivity at different wavelengths was extracted. The three bottom curves correspond to the oxidized TiO$_2$ device. **b**) Responsivity of the device at 405 nm as a function of incident power for the pristine device (top curve) and after consecutive heating cycles. **c**) Responsivity cut-off wavelength (top) and responsivity–power exponent (bottom) as a function of the heating cycle.

Figure 7b shows the evolution of the responsivity at 405 nm as a function of the incident optical power. In a photodetector, the responsivity at a certain wavelength $R(\lambda)$ versus the incident optical power $P$ can be described by a power law according to the formula:

$$R(\lambda) = P^{\alpha} \tag{1}$$

where $\alpha$ is a dimensionless exponent, which assumes a value of −0.5 in the case of a photodetector dominated by bimolecular recombination between photoexcited carriers or at high injection levels (band-to-band) and 0 in the case of monomolecular recombination or low injection levels (trap-assisted) [43,44]. The log–log plot shown in Figure 6b reveals that our device is characterized by a negative $\alpha$ for all the heating cycles since the slope of each curve in the figure gives directly the exponent $\alpha$. Figure 7c shows the extracted cut-off wavelength and the exponent $\alpha$ of the device as a function of the heating cycle. The evolution of the cut-off wavelength (here defined again as the largest wavelength for which the photodetector shows a response higher than the noise level $2 \times 10^{-6}$ A/W) is consistent with the predicted bandgap evolution shown in Figure 4a. As can be seen the exponent $\alpha$ starts from a value of −0.42 which indicates that the pristine $TiS_3$ device is mostly dominated by band-to-band recombination (0 heating cycle). After the first heating cycle the exponent decrease to a value of −0.7 that is the minimum value observed for $\alpha$. In the subsequent heating cycles (2–6) we observe a gradual increase in the value of the exponent $\alpha$ toward 0, signifying an increase in the density of traps and/or an effect on the lower effective light injection levels due to the lower absorption in the visible range of the spectrum by the $TiO_2$ (see Section S1 of the Supporting Information). The exponent $\alpha$ saturates around −0.25, a value larger than the starting value of −0.4, which indicates that the final $TiO_2$ material contains a larger trap density than the starting $TiS_3$ material that favors monomolecular recombination between photoexcited carriers [45,46].

## 4. Conclusions

In conclusion, we studied the thermal oxidation of $TiS_3$ nanoribbons with optical spectroscopy showing that this material can be gradually converted to anatase $TiO_2$ in a controlled way. We built photodetectors based on single $TiS_3$ nanoribbons and we tuned their cut-off wavelength by gradually oxidizing the nanoribbons. We observe that the oxidation also induces a decrease in the responsivity, which could have practical implications for the applicability of the $TiO_{2-x}S_x$, as a larger detector area would be required to obtain the same photocurrent. Ab-initio calculations of the band-structure of the materials are in agreement with the experiments. The shift of the cut-off wavelength in our photodetectors with a simple annealing step, demonstrates the ability to tune on-demand the bandgap of the $TiO_{2-x}S_x$ for novel applications.

**Supplementary Materials:** The following are available online at www.mdpi.com/xxx/s1, Section S1: additional optical characterization of the oxidation of $TiS_3$ nanoribbons (includes Figures S1, S2 and S3). Section S2: $TiS_3$ photodetector annealed in vacuum (includes Figure S4). Section S3: $TiS_3$ photodetector annealed in air (includes Figures S5, S6 and S7). Section S4: Stability of $TiS_3$ during Raman spectroscopy (includes Figures S8 and S9).

**Funding:** This research was funded by European Commission under the Graphene Flagship, grant number CNECTICT-604391, the Netherlands Organisation for Scientific Research (NWO) grant number 680-50-1515, the European Union's Horizon 2020 Marie Skłodowska-Curie grant number 793318, the Spanish Ministerio de Economia y Competitividad grant number FIS2016-79464-P, the Grupo Consolidado UPV/EHU del Gobierno Vasco grant number IT578-13, MINECO-FEDER grant number MAT2015-65203-R, the Elemental Strategy Initiative conducted by the MEXT, Japan and the CREST grant number JPMJCR15F3.

**Acknowledgements:** AC-G acknowledges funding from the European Commission under the Graphene Flagship, contract CNECTICT-604391. RF acknowledges support from the Netherlands Organisation for Scientific Research (NWO) through the research program Rubicon with project number 680-50-1515. RB acknowledges funding from the European Union's Horizon 2020 research and innovation program under the Marie Skłodowska-Curie grant agreement No. 793318. RB and RDA acknowledge financial support by SElecT-DFT (Grant No. FIS2016-79464-P) of the Spanish Ministerio de Economia y Competitividad through the Agencia Estatal de Investigacion and the Fondo Europeo de Desarrollo Regional and Grupo Consolidado UPV/EHU del Gobierno Vasco (IT578-13). MIRE Group acknowledges funding from MINECO-FEDER through the project MAT2015-65203-R. KW and TT acknowledge the support of the Elemental Strategy Initiative conducted by the MEXT, Japan and the CREST (JPMJCR15F3), JST.

**Conflicts of Interest**: The authors declare no competing financial interests.

# Supporting Information: Tunable photodetectors via *in situ* thermal conversion of TiS$_3$ to TiO$_2$

Foad Ghasemi[1,2,+], Riccardo Frisenda*[3,+], Eduardo Flores[4], Nikos Papadopoulos[5], Robert Biele[6,7], David Perez de Lara[1], Herre van der Zant[5], Kenji Watanabe[8], Takashi Taniguchi[8], Roberto D'Agosta[6,9], Jose R. Ares[4], Carlos Sánchez[4,10], Isabel J. Ferrer[4,10] and Andres Castellanos-Gomez*[3]

[1] Instituto Madrileño de Estudios Avanzados en Nanociencia (IMDEA-Nanociencia), Campus de Cantoblanco, E-28049 Madrid, Spain.

[2] Nanoscale Physics Device Lab (NPDL), Department of Physics, University of Kurdistan, 66177-15175 Sanandaj, Iran.

[3] Materials Science Factory, Instituto de Ciencia de Materiales de Madrid (ICMM-CSIC), E-28049, Madrid, Spain.

[4] Materials of Interest in Renewable Energies Group (MIRE Group), Dpto. de Física de Materiales, Universidad Autónoma de Madrid, UAM, Campus de Cantoblanco, E-28049 Madrid, Spain.

[5] Kavli Institute of Nanoscience, Delft University of Technology, Lorentzweg 1, Delft 2628 CJ, The Netherlands.

[6] Nano-Bio Spectroscopy Group and European Theoretical Spectroscopy Facility (ETSF), Universidad del País Vasco CFM CSIC-UPV/EHU-MPC & DIPC, Av.Tolosa 72 ,20018, San Sebastián, Spain.

[7] Institute for Materials Science and Max Bergmann Center of Biomaterials, TU Dresden, 01062 Dresden, Germany.

[8] National Institute for Materials Science, Namiki 1-1, Tsukuba, Ibaraki 305-0044, Japan.

[9] IKERBASQUE, Basque Foundation for Science, 48013 Bilbao, Spain.

[10] Instituto Nicolás Cabrera, Universidad Autónoma de Madrid, UAM, Campus de Cantoblanco, E-28049 Madrid, Spain.

*E-mail: riccardo.frisenda@csic.es, andres.castellanos@csic.es.

[+] **These two authors contributed equally to this work.**

## Section S1 – Additional optical characterization of the oxidation of TiS₃ nanoribbons

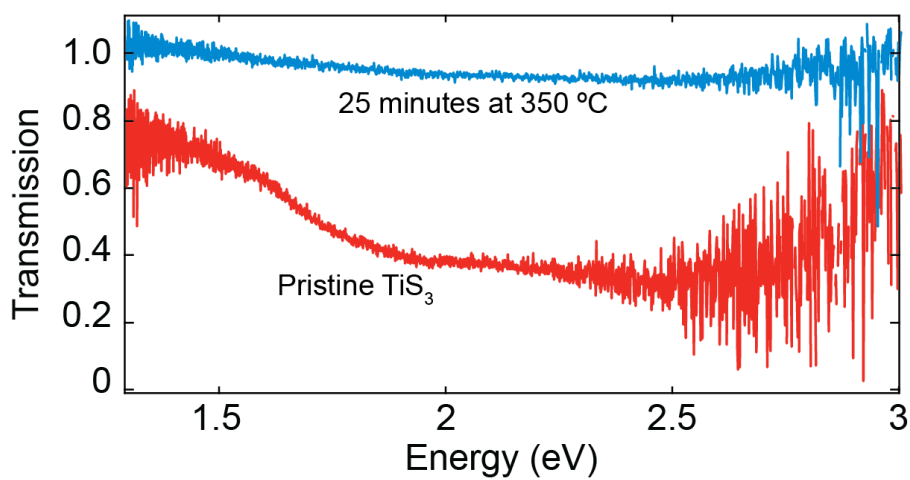

**Figure S1**: Energy resolved transmission of TiS₃ and TiO₂.

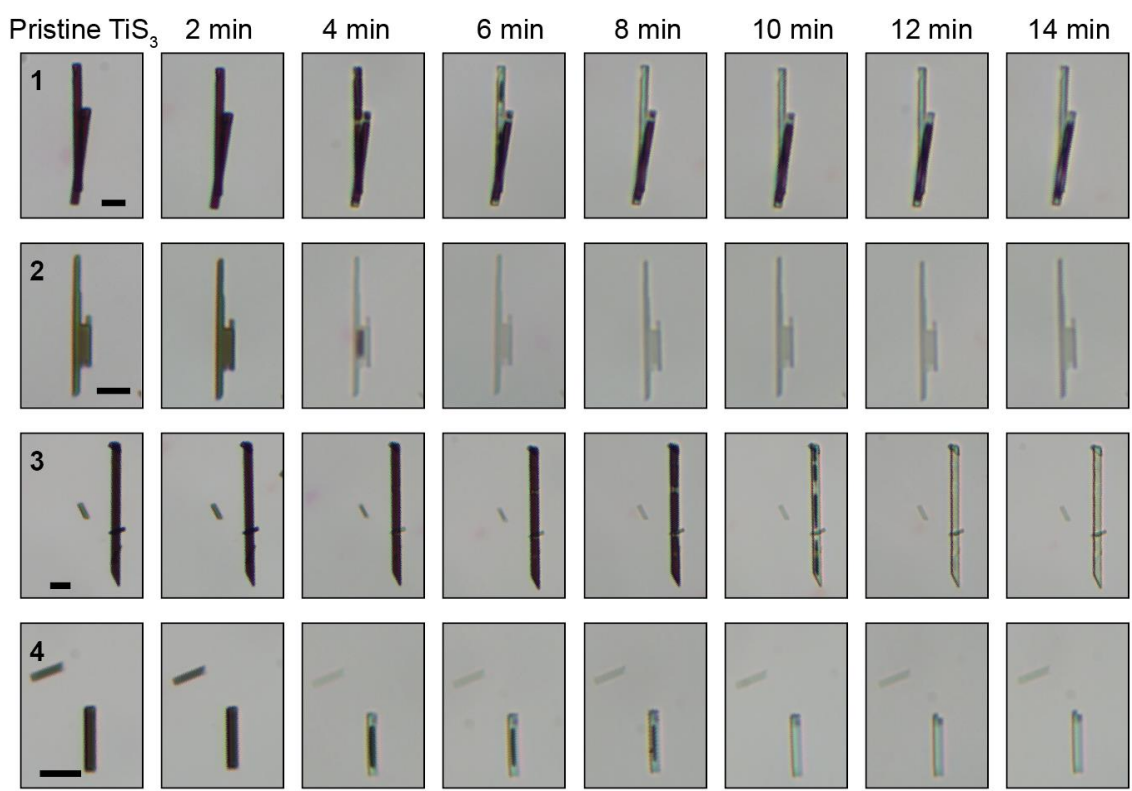

**Figure S2**: Optical image of individual TiS₃ nanoribbons transferred onto an ITO/glass substrate. The different rows show various nanoribbons identified in the same substrate. The leftmost picture of each row has been recorded on the pristine TiS₃, before heating the substrate. The subsequent images have been recorded after heating the sample to 320 °C for the time indicated in each column. The black scale bars in the left-most pictures correspond to 10 µm.

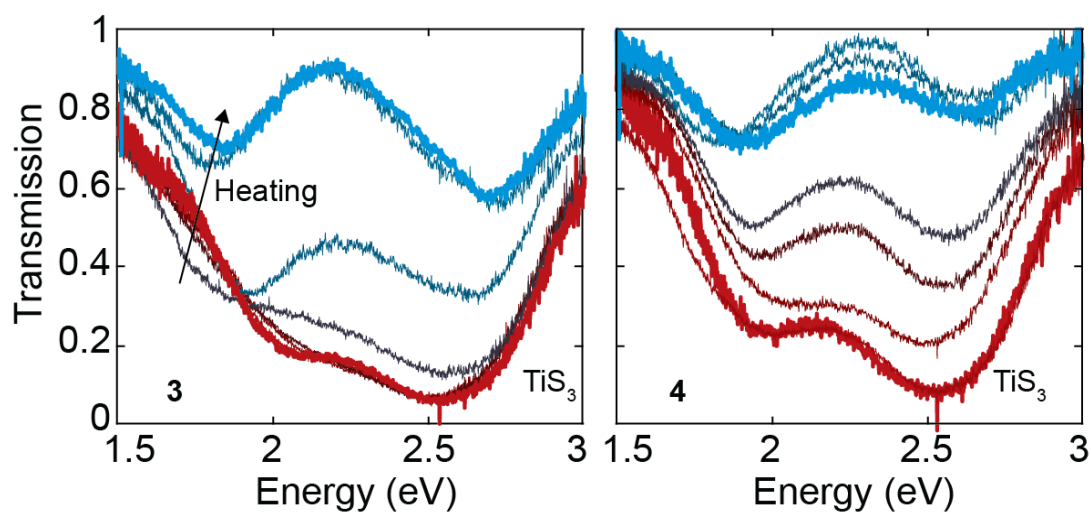

**Figure S3**: Energy resolved transmission of ribbons **3** and **4** from Fig. S2. The curves drawn with thick lines correspond to the initial (red) and final (blue) state of the ribbon under study.

### Section S2 – TiS$_3$ photodetector annealed in vacuum

As a control experiment, to investigate the influence of the temperature in absence of oxygen, we studied a TiS$_3$ photodetector that we heated above 320 °C in vacuum. Figure S4a shows the responsivity of the device in its pristine state (red curve) and after 10, 20 and 30 minutes at 320 °C in vacuum. The shape of the responsivity (and the power exponent in panel b) is preserved and the device shows an almost flat response from less than 400 nm to more than 700 nm (as can be also seen in panel c), typical of TiS$_3$ photodetectors.

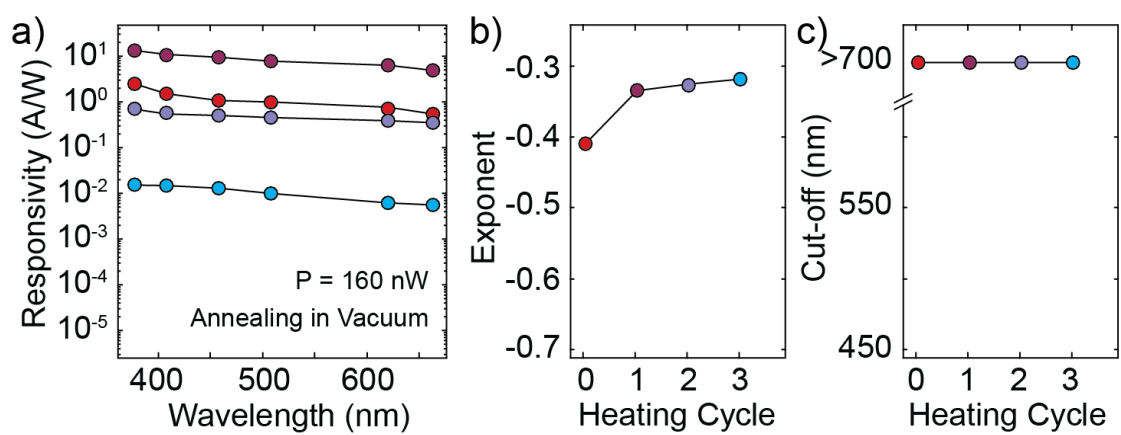

**Figure S4**: a) Responsivity of the device as a function of wavelength. The pristine TiS$_3$ device (top) was heated at 320 °C in vacuum in steps of 10 minutes and after each step

the responsivity was measured. Note that the dark current decreases for the successive heating cycles from 70 µA (pristine) to 20, 1.1 and 0.025 µA for the 1st, 2nd and 3rd heating cycle respectively. b) Responsivity exponent as a function of heating cycle. c) Cut-off wavelength as a function of heating cycle.

**Section S3 – TiS$_3$ photodetectors annealed in air**

In our study we characterized a total of 12 TiS$_3$ photodetecting devices, 11 devices have been heated above 320 °C in air (devices 1-11) and 1 device has been heated above 320 °C in vacuum as a control experiment. When heating up the device in air we managed to successfully convert the TiS$_3$ photodetector in a TiO$_2$ photodetector without losing the functionality in 4 devices, corresponding to a success rate of 36%. Figure S5a shows the responsivity as a function of wavelength of devices 1-11 each in its pristine state and Figure S5b shows the responsivity of devices 1-4, which have been successfully converted to TiO$_2$ photodetectors, in their initial and final states.

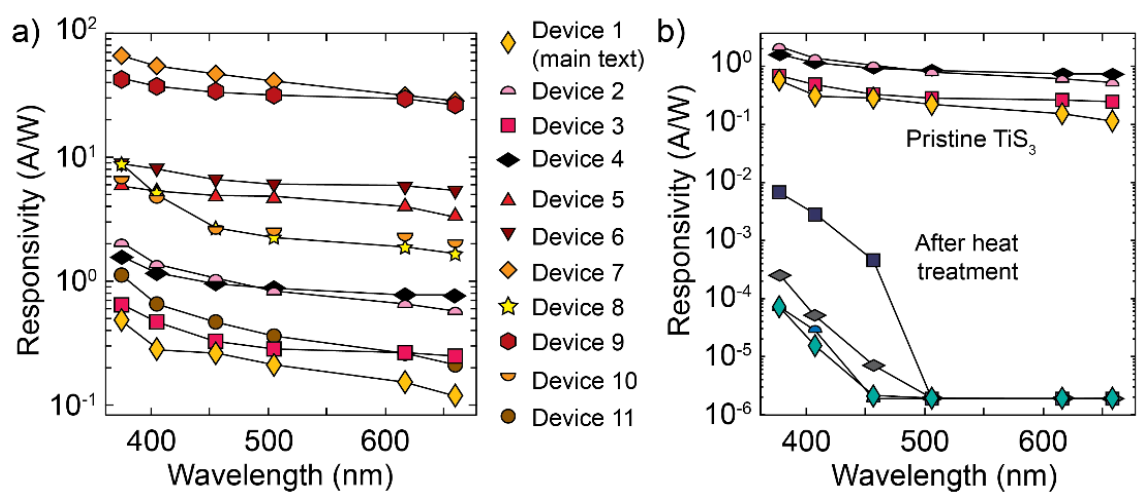

**Figure S5**: a) Responsivity as a function of wavelength of the pristine TiS$_3$ devices investigated in this work. b) Responsivity of the TiS$_3$ devices which have been successfully converted in TiO$_2$ photodetectors by heating at 320 °C. The four bottom curves correspond to the oxidized TiO$_2$ devices.

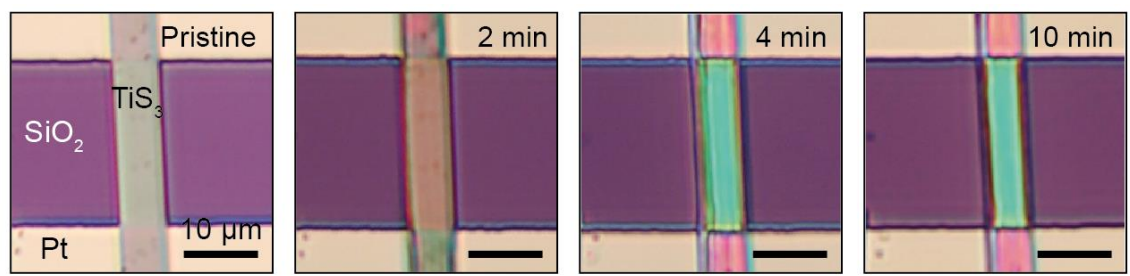

**Figure S6**: Optical microscope image of a TiS$_3$ photodetector (device 1) just after fabrication (left panel) and at different steps of the heating process.

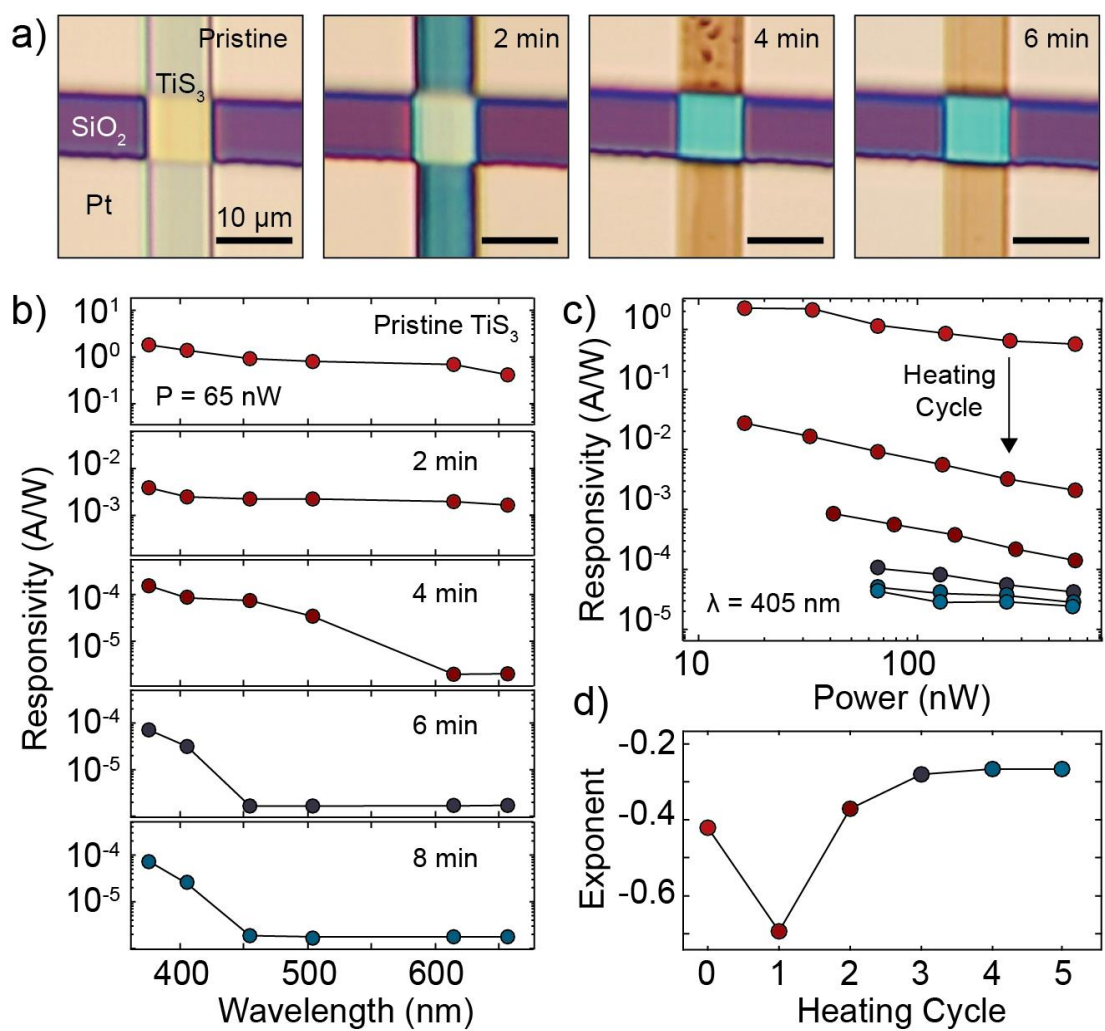

**Figure S7**: a) Optical microscope image of a TiS$_3$ photodetector (device 2) just after fabrication (left panel) and at different steps of the heating process. b) Responsivity of the device as a function of wavelength for different heating cycles. The pristine TiS$_3$ device (top) was heated at 320 °C in steps of 2 minutes and after each step the responsivity at different wavelengths was extracted. The two bottom curves correspond to the oxidized TiO$_2$ device. c) Responsivity of the device at 405 nm as a function of incident power for the pristine device (top curve) and after consecutive heating cycles. c) Responsivity-power law exponent as a function of heating cycle.

### Section S4 –Stability of TiS$_3$ during Raman spectroscopy

We performed measurements of TiS$_3$ ribbons at various incident optical power densities and integration times to probe the stability of TiS$_3$ during the Raman spectroscopy experiments. The results, which are collected in Figs. S8 and S9, indicate that TiS$_3$ ribbons undergo photooxidation at an incident power of 10 mW (spot size ~2 um$^2$, power density 5 mW/μm$^2$). In the case of lower densities we do not observe any degradation or photooxidation for exposition times as large as 60 s.

Figure S8a shows a semilogarithmic representation of the Raman spectra of a $TiS_3$ ribbon recorded for different incident powers (in sequence 0.2 mW, 2 mW, 5 mW, 10 mW, 20 mW, 0.2 mW) and different integration times. In all the three panels one can see that the spectra recorded at the lowest excitation power of 0.2 mW in the pristine state and after the application of the larger powers are different. While the pristine spectrum (red) shows only the peaks due to $TiS_3$, the final spectrum shows an additional peak at 142 cm$^{-1}$, due to $TiO_2$, independent on the integration time. This indicates that the laser at 532 nm used in the Raman experiments can oxidize the $TiS_3$ ribbons. Figure S8b shows the spectra recorded with integration time 10 s with a vertical offset added for clarity. From these spectra it is clear that the additional peak at 142 cm$^{-1}$ appears during the measurement at 10 mW. To quantify this phenomenon we perform a fit of the peaks at 142 cm$^{-1}$ (due to $TiO_2$) and at 301 cm$^{-1}$ (due to $TiS_3$) indicated by the shaded areas in panel b. Figure S8c shows the ratio between the areas of the peaks at 301 cm$^{-1}$ and 142 cm$^{-1}$ as a function of the incident power. By inspecting the plot we see that the ratio between the peaks is constant for powers as large as 5 mW (power density 2.5 mW/µm$^2$) and that rapidly decreases to zero (in an irreversible way) at a power of 10 mW (power density 5 mW/µm$^2$). In the case of an incident power equal or lower than 5 mW we do not observe any laser induced oxidation for integration times as long as 60 s as can be seen in Figure S9.

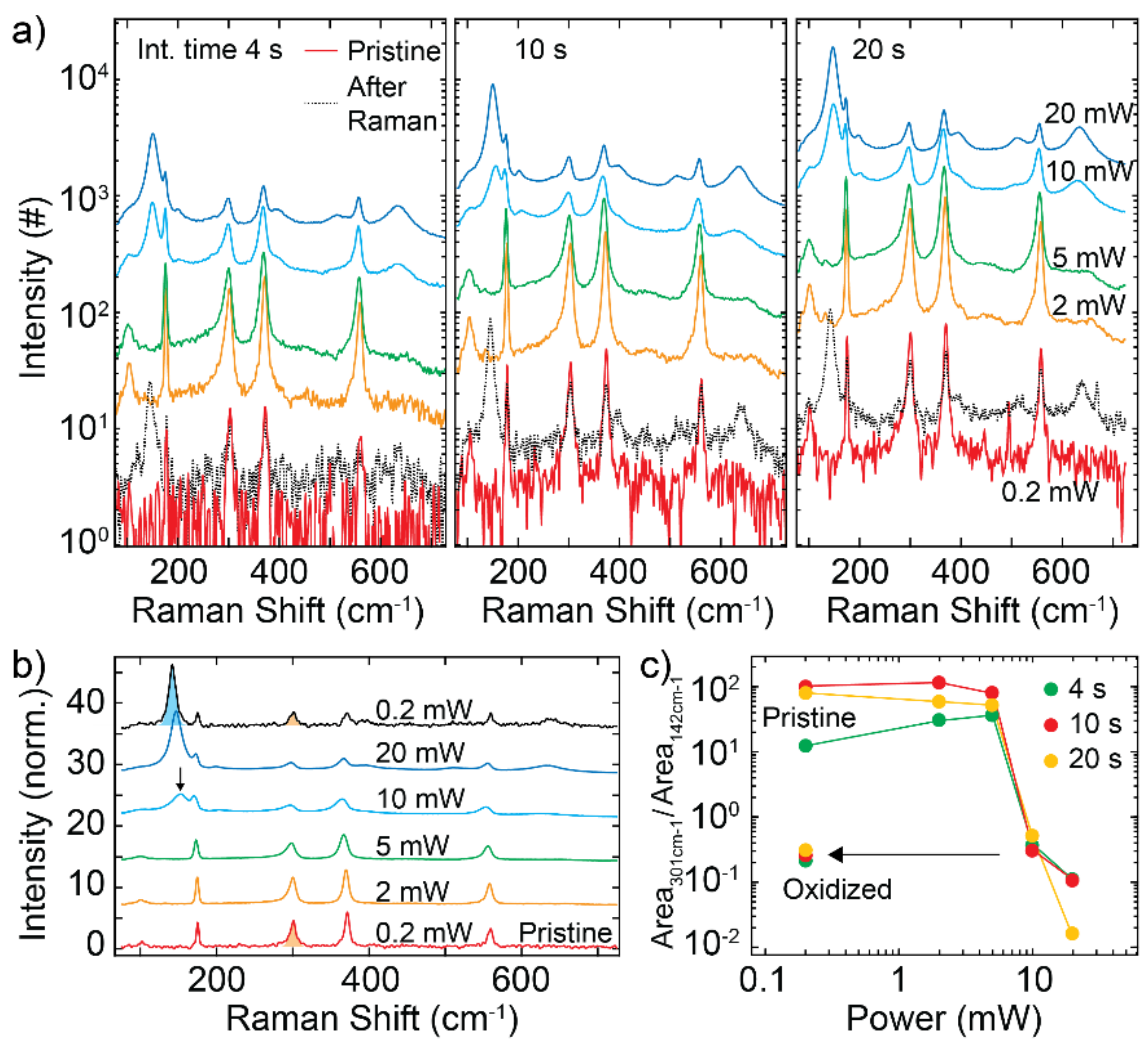

**Figure S8**: a) Raman spectroscopy of a TiS$_3$ ribbon as a function of illumination power (curves with different colours) and integration time. For each integration time (4 s, 10 s and 20 s) we select a different position in the ribbon to probe the pristine material. b) Raman spectra recorded with integration time 10 s at different incident powers. Each spectrum is normalized by the incident power. A vertical offset has been introduced for clarity. c) Ratio between the areas of the peaks at 301 cm$^{-1}$ (related to TiS$_3$) and at 142 cm$^{-1}$ (TiO$_2$) as a function of power.

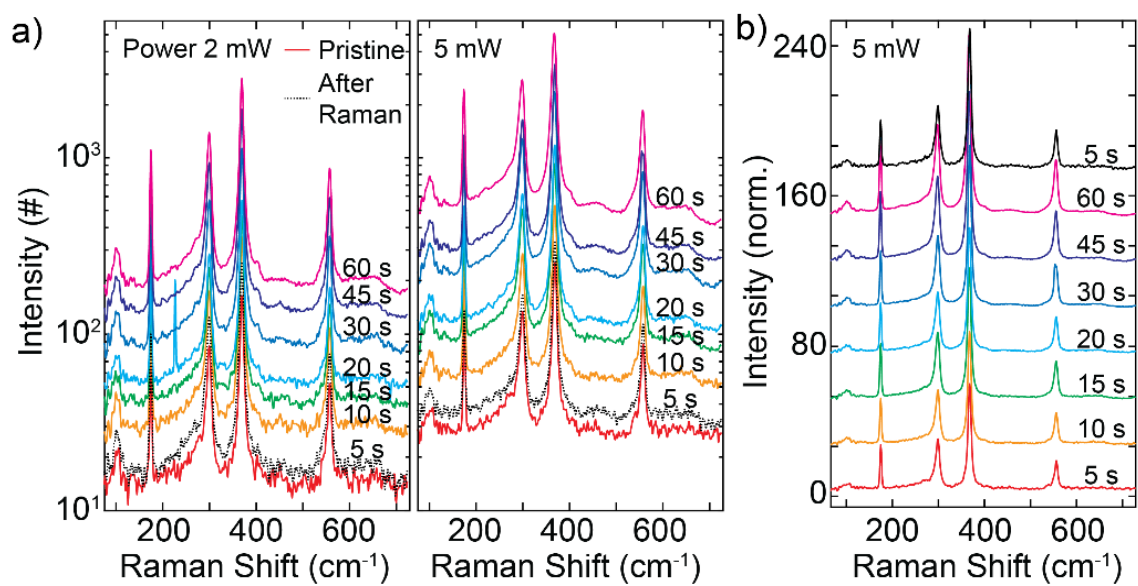

**Figure S9**: a) Raman spectroscopy of a TiS$_3$ ribbon as a function of integration time (curves with different colours) for different powers. For each power we select a different position in the ribbon to probe the pristine material. b) Raman spectra recorded with incident power 5 mW time at different integration times. Each spectrum is normalized by the integration time. A vertical offset has been introduced for clarity.